\def\preprint{0}                
\def\preprint{1}                

\if\preprint1
        \documentclass{article}
        \usepackage{mn2e}
        \usepackage{astrop-bib}
        \usepackage{times}
        \usepackage{graphicx}
        \usepackage{calc}

\else
        \documentclass{article}
        \usepackage[referee]{mn2e}
        \usepackage{astrop-bib}
        \usepackage{times}
        \newcommand{\includegraphics}[2][]{\relax} 
\fi

\newcommand{\aCenA}{\mbox{$\alpha$~Cen~A}}
\newcommand{\eBoo}{\mbox{$\eta$~Boo}}
\newcommand{\aCen}{\mbox{$\alpha$~Cen}}
\newcommand{\secref}[1]{Section~\ref{#1}}
\newcommand{\Secref}[1]{\secref{#1}}
\newcommand{\figref}[1]{Figure~\ref{#1}}
\newcommand{\Figref}[1]{\figref{#1}}
\newcommand{\tabref}[1]{Table~\ref{#1}}
\newcommand{\eqref}[1]{Equation~\ref{#1}}
\newcommand{\nn}{\nonumber}
\newcommand{\muHz}{\mbox{$\mu$Hz}}
\newcommand{\Dnu}[1]{\Delta \nu_{#1}}
\newcommand{\half}{{\textstyle\frac{1}{2}}}
\ifnfsstwo

\fi

\ifnfssone
  \newmathalphabet{\mathit}
    \addtoversion{normal}{\mathit}{cmr}{m}{it}
    \addtoversion{bold}{\mathit}{cmr}{bx}{it}

\fi

\ifoldfss

\fi

\loadboldmathitalic
\loadboldgreek

\let\epsilon\varepsilon

\title[Oscillations and granulation in \aCenA]{A search for solar-like
oscillations and granulation in $\balpha$~Cen~A}

\author[H.~Kjeldsen et al.]
       {H.~Kjeldsen,$^{1,2}$
        T.~R.~Bedding,$^3$
        S. Frandsen$^2$ and T.~H.~Dall$^2$\\
        $^1$Teoretisk Astrofysik Center, Danmarks Grundforskningsfond
		{\tt (hans@obs.aau.dk)}\\
        $^2$Institute of Physics and Astronomy, Aarhus University, DK-8000
                Aarhus C, Denmark\\
        $^3$Chatterton Astronomy Department, School of Physics, University
                of Sydney 2006, Australia {\tt
                (bedding@physics.usyd.edu.au)}
}
\pubyear{1998}

\begin{document}

\maketitle

\begin{abstract}
We report the most sensitive search yet made for solar-like oscillations.
We observed the star \aCenA{} in Balmer-line equivalent widths over six
nights with both the 3.9\,m Anglo-Australian Telescope and the ESO 3.6\,m
telescope.  We set an upper limit on oscillations of 1.4 times solar and
found tentative evidence for $p$-mode oscillations.  We also found a power
excess at low frequencies which has the same slope and strength as the
power seen from granulation in the Sun.  We therefore suggest that we have
made the first detection of temporal fluctuations due to granulation in a
solar-like star.
\end{abstract}


\section{Introduction}

Many attempts have been made to detect stellar analogues of the solar
five-minute oscillations.  As with helioseismology, it is hoped that the
measurement of oscillation frequencies in other stars will place important
constraints on stellar model parameters and provide a strong test of
evolutionary theory.  However, despite several claims in the literature, it
is fair to say that there has been no unambiguous detection of solar-like
oscillations in any star except the Sun (see reviews by \citebare{B+G94};
\citebare{K+B95}; \citebare{G+S96}; \citebare{HJLaB96}; \citebare{B+K98}).

Most searches for stellar oscillations have used one of two methods:
(i)~high-resolution spectroscopy, which aims to detect periodic Doppler
shifts in spectral features, or (ii)~differential CCD photometry, which has
been used to search for fluctuations in the integrated luminosity of stars
in the open cluster M67 \cite{GBK93}.

We have been using a new method to search for solar-like oscillations that
involves measuring temperature changes via their effect on the equivalent
widths (EWs) of the Balmer hydrogen lines.  We found strong evidence for
solar-like oscillations in the G~subgiant \eBoo\ \cite{KBV95,B+K95}, with
frequency splittings that were later found to agree with theoretical models
(\citeabbare{ChDBH95}{ChDBK95}; \citebare{G+D96}).  Since then, the
improved luminosity estimate for \eBoo{} from Hipparcos measurements has
given even better agreement \cite{BKChD98}.  However, a search for velocity
oscillations in \eBoo{} by \citeone{BKK97} failed to detect a signal,
setting limits at a level below the value expected on the basis of the
\citename{KBV95} result.  More recently, Brown et al.\ (in preparation)
have obtained a larger set of Doppler observations with lower noise which
also fail to show convincing evidence for oscillations.



Meanwhile, the equivalent-width method has now been used to detect
oscillations in the spatially resolved Sun \cite{KHB98}, in the
$\delta$~Scuti variable FG Vir \cite{VKB98} and in the rapidly oscillating
Ap star $\alpha$~Cir \cite{BVB99}.  For the Sun it has been shown directly
that the EW variations of Balmer lines arise because the line intensity
oscillates much less than the continuum intensity \cite{RHD91,KHB98}.

We chose \eBoo\ as the first solar-like target for the EW method because it
was expected to have an oscillation amplitude about five times greater than
the Sun.  This turned out to be the case (assuming the detection is real).
We then turned to \aCenA, a more challenging target because of its smaller
expected oscillation amplitude (comparable to solar; \citebare{BKR96}).
Being a near twin of the Sun and extremely nearby, this star is an obvious
target to search for oscillations \citeeg{BCW94}.  Previous attempts using
Doppler methods were reviewed by \citeone{K+B95}.  They include two claimed
detections at amplitudes 4--6 times greater than solar \cite{GGF86,PBvH92}
and two negative results at amplitudes about 2--3 times solar
\cite{B+G90,E+C95}.

Here, we report observations of \aCenA{} in Balmer-line equivalent widths
which set an upper limit of only 1.4 times solar.  We find tentative
evidence for $p$-mode oscillations.  We also find strong evidence for
stellar granulation in the power spectrum, at a level consistent with that
expected on the basis of solar observations.

\section{Observations}

We observed \aCenA\ over six nights in April 1995 from two sites: 
\begin{itemize}

\item At Siding Spring Observatory in Australia, we used the 3.9-metre
Anglo-Australian Telescope (AAT) with the coud\'e echelle spectrograph
(UCLES).  We recorded three orders centred at H$\alpha$ and three orders at
H$\beta$, which was possible thanks to the flexibility of the CCD
controller.  The weather was about 85\% clear but transparency was
variable.  The exposure time was typically about 30\,s, with a deadtime of
23\,s between exposures.  This relatively long exposure time for such a
bright target was achieved by spreading the light along the spectrograph
slit by (i)~defocussing the telescope and using a wide slit (5--10\,arcsec)
and (ii)~trailing the star backwards and forwards along the slit with a
peak-to-peak amplitude of about 2\,arcsec and a period of about 4\,s.  The
slit length was 14\,arcsec.  Each spectrum typically produced about
$9.0\times10^7$\,photons/\AA{} in the continuum near H$\alpha$ and about
$3.5\times10^7$\,photons/\AA{} near H$\beta$.  

\item At the European Southern Observatory on La Silla in Chile, we used
the ESO 3.6-metre telescope with the Cassegrain echelle spectrograph
(CASPEC).  We recorded three orders centred at H$\alpha$.  The weather was
100\% clear.  The exposure time was typically about 10\,s, with a deadtime
of 12\,s between exposures.  The slit was 7\,arcsec wide and 10\,arcsec
long, with slightly less defocus than at the AAT and no trailing.  Each
spectrum typically produced about $5.5\times10^7$\,photons/\AA{} in the
continuum near H$\alpha$.

\end{itemize}

The B component of the \aCen{} system, which is 1.3 magnitudes fainter than
the A component, had a separation of 17.3\,arcsec at the time of our
observations, so there should not be any contamination.  We nevertheless
kept the spectrograph slit aligned with the position angle of the binary
system so that any light from the B component would be spatially separated
on the detector.

In addition to our primary target of \aCenA, we also observed Procyon for
one hour at the start of each night, plus 5 hours at both sites at the start
of the fifth night.  Finally, we observed the solar spectrum via the
daytime sky (about 18\,hr at AAT and 14\,hr at ESO, spread over the six
days).  For both Procyon and the Sun, this amount of data turned out not to
be sufficient to detect oscillations, but we did find evidence for
granulation power (see \secref{sec.granulation}).

\section{Data processing}

Here we describe the steps involved in processing the data.
\Secref{sec.step.ew} is specific to the equivalent-width method, while the
other steps could apply, at least in part, to other types of observations
(Doppler shift or photometry).

\subsection{Preliminary reduction}

        \begin{enumerate}  

        \item Correction of each CCD frame for bias by subtracting an
	average bias frame and then subtracting a constant that was
	measured from the overscan region of the CCD frame.

        \item Correction for CCD non-linearity.  Measuring oscillations at
        the ppm level requires that the detector be linear to the level of
        one part in 1000 or better.  This cannot be taken for granted and
        our tests of different CCDs and controllers often reveal deviations
        from linearity of up to a few per cent.  Unless correction is made
        for these effects, the extra noise will destroy any possibility of
        detecting oscillations.

	Both CCDs were Tektronix 1024$\times$1024 chips, used at a
	conversion factor of about 12 electrons per ADU\@.  The onset of
	saturation was about 450\,000 electrons per pixel for the AAT and
	250\,000 for ESO\@.  {}From our measurements, the linearity
	corrections for both were well approximated by the relation:
	\begin{eqnarray}
	\mbox{ADU}_{\rm obs} / \mbox{ADU}_{\rm true}
		 & = & 1 + \alpha\; \mbox{ADU}_{\rm obs}  \nn\\
		 &   & - \exp(\mbox{ADU}_{\rm obs}/\beta + \gamma),
	\end{eqnarray}
	where $\mbox{ADU}_{\rm obs}$ and $\mbox{ADU}_{\rm true}$ are
	observed and true counts measured in ADU and ($\alpha$, $\beta$,
	$\gamma$) had values of ($-1.63\times10^{-7}$, $700$, $-59.33$) for
	the AAT and ($-2.35\times10^{-7}$, $700$, $-33.53$) for ESO\@.  The
	measurement error on this correction was one part in 3500.

        \item Correction for pixel-to-pixel variations in CCD sensitivity
        by dividing by an average flat-field exposure (all flat-field
        exposures were corrected for non-linearities before averaging).

        \item Subtraction of sky background, which was quite substantial
        during twilight.  The background was estimated from the regions at
        each end of the spectrograph slit, either side of the stellar
        spectrum in each echelle order.

        \item Extraction of one-dimensional spectra.  During this step, the
        seeing in each frame (i.e., the FWHM along the spectrograph slit)
        and the position of the star (i.e., light centroid along the slit)
        were recorded for use in decorrelation (see Sec.~\ref{sec.filter}
        below).

        \end{enumerate}

\subsection{Measuring equivalent widths}        \label{sec.step.ew}

Achieving high precision requires more than simply fitting a profile.  The
method described here was developed after trying several different
approaches.  By analogy with Str{\o}mgren H$\beta$ photometry, we calculated
the flux in three artificial filters, one centred on the line ($L$) and the
others on the continuum both redward ($R$) and blueward ($B$) of the line.
For each spectrum, the following steps were followed:
        \begin{enumerate}

        \item Placement of the three filters at their nominal wavelengths.

        \item Calculation of the three fluxes $B$, $L$ and
         $R$.\footnote{The flux in a filter is simply the total counts in
         the stellar spectrum after it has been multiplied by the filter
         function.}

        \item Adjustment of the slope of the spectrum so that $R$ and $B$
        were equal.  This was done by multiplying the spectrum by a linear
        ramp.

        \item Re-calculation of the filter fluxes and hence of the
        equivalent width: $W = (R-L)/R$.

        \item Repetition of steps (ii)--(iv) with the three filters at
        different positions.  Iteration to find the filter position that
        maximized the value of $W$.  The outputs were: $W$, the position of
        the line, the height of the continuum (from $R$), and slope of
        continuum.

        \item Repetition of steps (i)--(v) for four different filter
	widths.

        \end{enumerate} 
The result was four times series ($W_1$, $W_2$, $W_3$, $W_4$), one for each
filter width.

\subsection{Initial time series processing}

The quality of the data, as measured by the local scatter, varied
considerably from hour to hour and night to night.  The following procedure
was applied to each night of data separately.
        \begin{enumerate}

        \item Clipping of each time series to remove outlying points
        (4$\sigma$ clipping, where $\sigma$ is the local rms scatter).

        \item Calculation of weights for each time series.  This involved
        assigning a weight to each data point that was inversely
        proportional to the local variance ($\sigma^2$).

        \item Calculation of $\sigma_w$, the {\em weighted\/} rms scatter
        of each time series, using the weights just calculated.  This was
        used to select the best filter width, i.e., the one which minimized
        $\sigma_w$.  By using a weighted rms scatter, we do not give too
        much importance to the bad segments of the data.  In practice,
        rather than choosing one filter width, we used a weighted
        combination.  That is, we chose the powers $a,b,c,d$ to minimize
        the weighted scatter on the time series $W_1^a W_2^b W_3^c W_4^d$,
        where $a+b+c+d=1$.

	\end{enumerate}
This step and the subsequent ones rely on the fact that any oscillation
signal will be much smaller than the rms scatter in the time series.  Most
of the scatter is due to noise and any method of reducing the scatter
should be a good thing, although care must be taken not to destroy the
signal or to introduce a spurious signal.

\subsection{Decorrelation of time series}       \label{sec.filter}

As well as measuring the parameter which is expected to contain the
oscillation signal ($W$), we also monitored extra parameters.  The aim was
to correct for instrumental and other non-stellar effects.  For example, if
$W$ is correlated with the seeing, we would suspect some flaw in the
reduction procedure, since the stellar oscillation does not know what is
happening in the Earth's atmosphere.  By correlating measured equivalent
widths with seeing variations, one has a chance to remove the influence of
seeing simply by subtracting that part of the signal which correlates with
seeing.  This process of decorrelation, which can be repeated for other
parameters (total light level, position on detector, slope of continuum,
etc.), is very powerful but can also be quite dangerous if not done with
care (see \citebare{GBD91} for a fuller discussion).

Again, the process was applied to data from one night at a time.
Performing decorrelation over shorter intervals runs the risk of moving
power around and creating or destroying signal -- simulations were useful
to check these effects.

Prior to decorrelation, we high-pass filtered the time series to remove
low-frequency variations.  This was done by subtracting from the time
series a smoothed version that was produced by convolution with a
supergaussian envelope having a FWHM of 40 minutes.

The AAT spectra included a strong Fe\,{\sc i} line near H$\beta$ at
4383\,\AA{} and we used the EW of this line as a decorrelation parameter.
This line is temperature sensitive, with the opposite sign and similar
amplitude to the Balmer lines \cite{BKR96}, so it should contain
oscillation signal in anti-phase with the Balmer lines -- this was allowed
for in the analysis.  We found that decorrelating against the Fe\,{\sc i}
EW gave a significant reduction in the scatter of both the H$\alpha$ and
H$\beta$ EW time series.  Unfortunately, the ESO spectra (around H$\alpha$)
did not contain any strong iron lines.

\if\preprint1
\begin{figure*} 

\includegraphics[
width=0.9\textwidth, bb= 90 147 524 394]
 {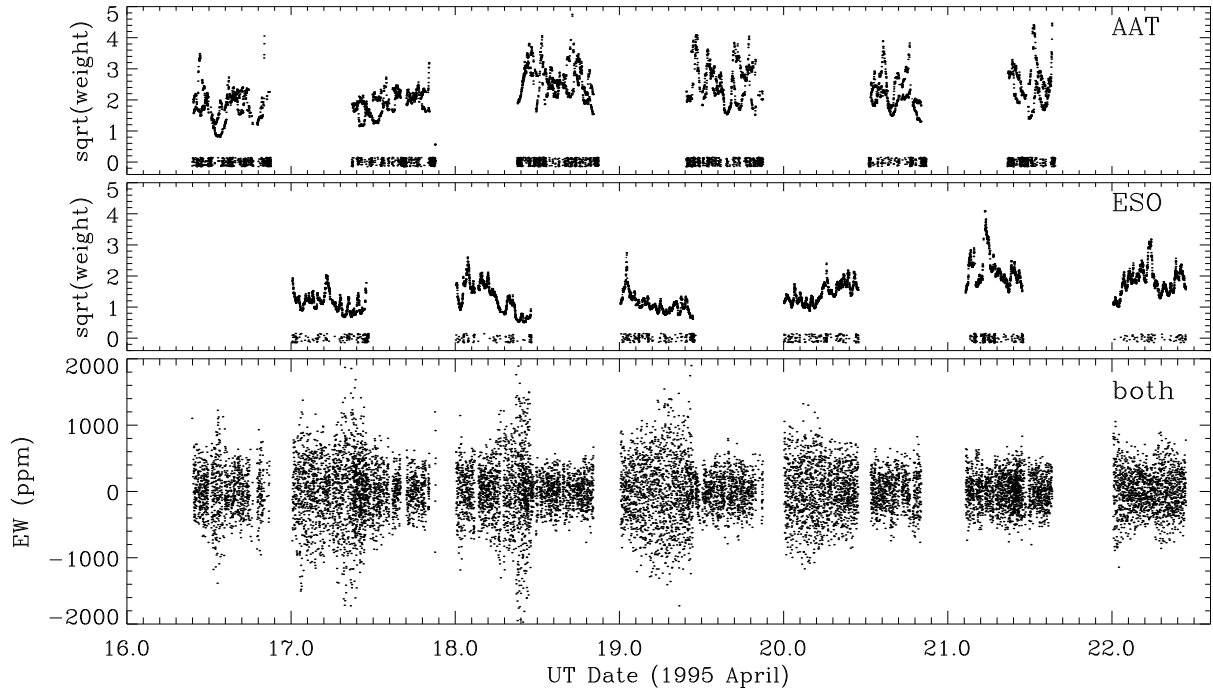}
\caption[]{\label{fig.weights} The final time series of EW observations of
\aCenA.  The upper panel shows the square root of statistical weights (the
inverse of the local rms scatter) in arbitrary units for all AAT
measurements (both H$\alpha$ and H$\beta$ are shown).  There are 8430
points, with 5773 having non-zero weight.  For points with zero weight, we
have added a random vertical scatter to make it easier to see their
distibution in time.  The middle panel shows the same for ESO data
(H$\alpha$ only); there are 8847 points, with 7999 having non-zero weight.
The bottom panel shows all EW measurements having non-zero weight (13772
points).  The peak-to-peak height of the individual modes we are trying to
detect corresponds approximately to the size of the dots in the figure.  }
\end{figure*}

  
\fi

\Figref{fig.weights} shows the final time series of EW measurements and
weights.  The data quality clearly varies significantly.  Almost one third
of the AAT spectra of \aCenA{} were discarded, although this is not quite
as bad as it sounds because many of these were taken well into morning
twilight.
Still, many night-time AAT spectra were given zero weight.  The reason for
these poor measurements is not clear, but it is probably related to
seeing-induced variations in the point-spread-function through the wide
slit (see below).  Those AAT measurements that are useful have much lower
scatter, and therefore higher weights, than ESO measurements because of the
availability of the Fe\,{\sc i} line for decorrelation.

\if\preprint1
\begin{figure*}
 \includegraphics[draft=false,width=15cm, bb=49 289 563 461] 
{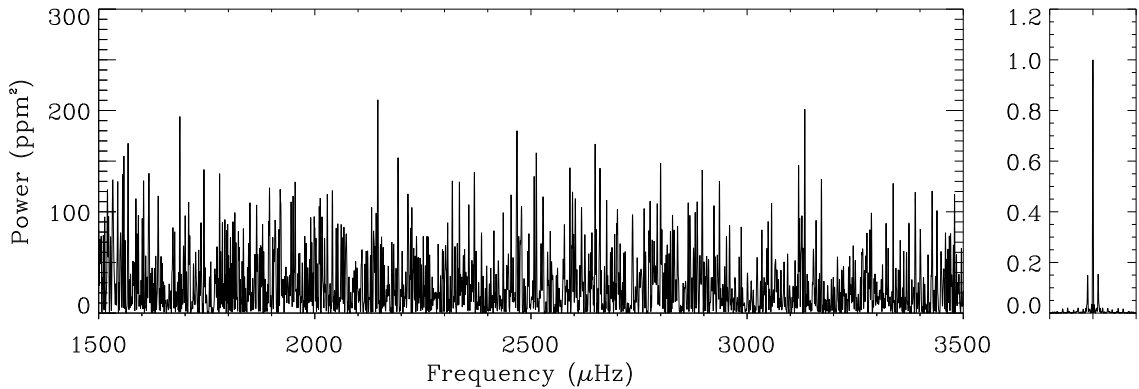}

\caption[]{\label{fig.acen}Power spectrum of equivalent-width observations
of \aCenA{} in the region where signal would be expected.  The right-hand
plot shows the spectral window, with the same horizontal scale, calculated
by taking the power spectrum of a sinusoid with an amplitude of 1\,ppm and
the same sampling and weights as the actual data. }
\end{figure*}

\fi

\section{Results and simulations}	\label{sec.results}

The amplitude spectrum of the EW time series was calculated as a weighted
least-squares fit of sinusoids \cite{FJK95,AKN98}, scaled so that a
sinusoid with an amplitude of 1\,ppm produces a peak of 1\,ppm.  The power
spectrum, the square of this amplitude spectrum, is shown in
\figref{fig.acen} for the region in which oscillations are expected.  No
obvious excess of power is seen.  We can fit a two-component noise model to
the whole power spectrum which consists of (i)~white (i.e., flat) noise at
a level of 4.3\,ppm in the amplitude spectrum, and (ii)~a non-white
component, which is discussed in \secref{sec.granulation}.

The total noise at 2.3\,mHz is 4.7\,ppm in the amplitude spectrum, while
that expected from photon statistics alone is 2.4\,ppm.  We are therefore a
factor of 1.8 away from the photon noise, with the extra noise presumably
coming from instrumental and/or atmospheric effects.  The use of a wide
slit, while advantageous in terms of exposure times and duty cycle, led to
seeing-induced changes in the point-spread-function which may be the cause.


\subsection{Simulations}

We expect oscillations in \aCenA{} to produce a regular series of peaks in
the power spectrum, with amplitudes modulated by a broad envelope centred
at about 2.3\,mHz \cite{K+B95}.  The average amplitude of oscillation modes
near the peak of this envelope, as measured in Balmer line EW, is expected
to be about 8\,ppm, which is 1.4 times that for the Sun \cite{BKR96}.  To
investigate whether an oscillation signal may be present in our data, we
have generated simulated time series consisting of artificial signal plus
noise.  Each simulated series had exactly the same sampling function and
allocation of statistical weights as the real data.  The injected signal
contained sinusoids at the frequencies calculated by \citeone{ECD92} for
modes with $\ell = 0$ to $3$, modulated by a broad solar-like envelope
centred at 2.3\,mHz and with a central height of 8\,ppm.  In each
simulation, the phases of the oscillation modes were chosen at random and
the amplitudes were randomized about their average values.  All these
characteristics were chosen to imitate as closely as possible the
stochastic nature of oscillations in the Sun.  Before calculating each
power spectrum, we added normally-distributed noise to the time series, so
as to produce a noise level in the amplitude spectrum of 4.7\,ppm
(consistent with the actual data).

\if\preprint1
\begin{figure*} 

\includegraphics[width=13cm, bb=74 56 503 648]
 {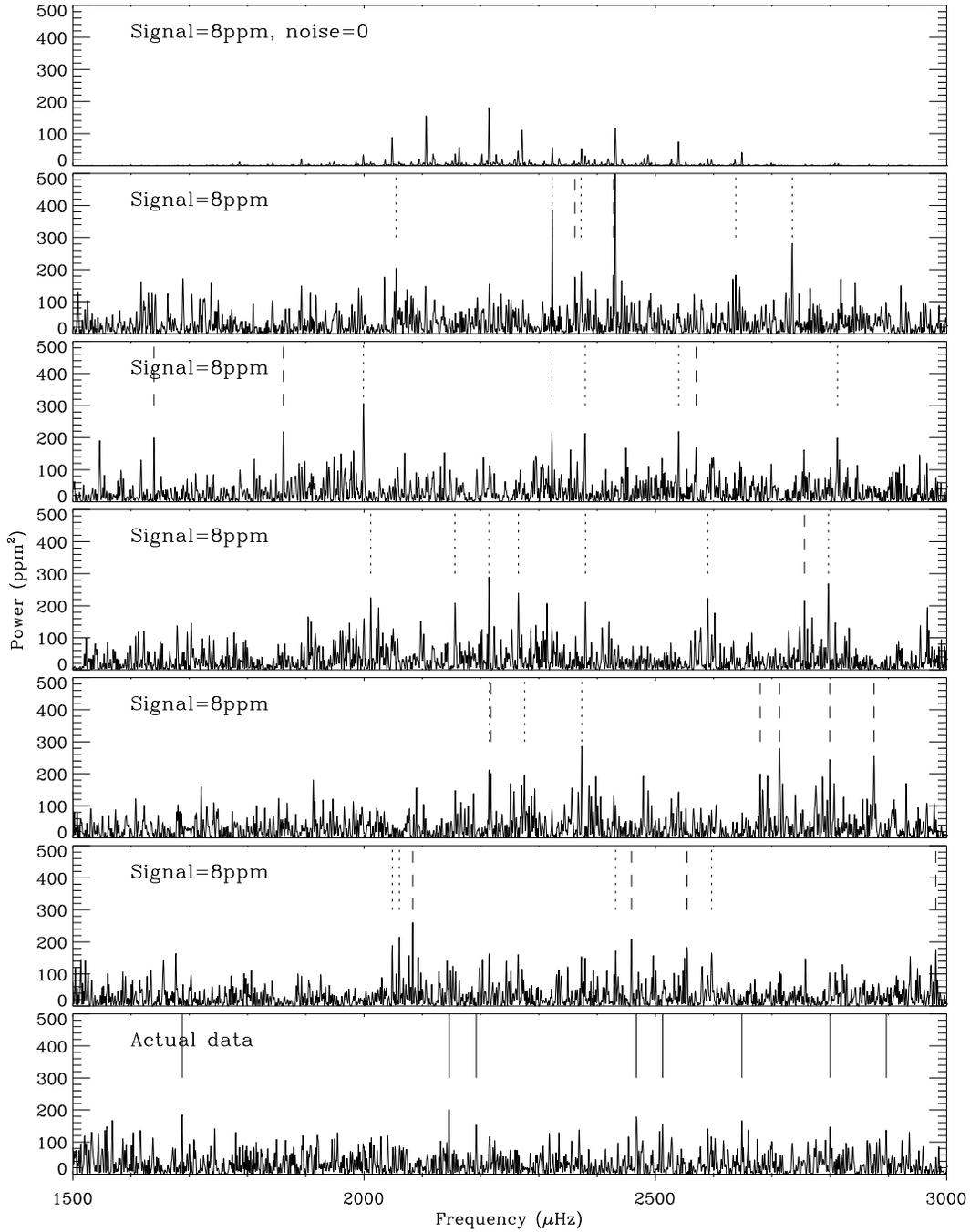}
\caption[]{\label{fig.sim}Simulated power spectra for \aCenA, using the
same sampling times and data weights as the actual observations.  The top
panel shows a simulation without noise, while the next five panels show
simulations with noise.  The eight strongest peaks are marked by vertical
lines, with dotted lines indicating those coinciding with input frequencies
or their 1/day aliases, and dashed lines showing noise peaks (see also
\figref{fig.echelle}).  The bottom panel shows the actual data from \aCenA,
with the eight strongest peaks being marked (see \tabref{tab.freqs}).  }
\end{figure*}

\fi

Some results are shown in \figref{fig.sim}.  The top panel shows a
simulation without any added noise; the randomization of mode amplitudes
within the broad envelope is clear.  The next five panels show simulations
with noise included (each with different randomization of noise, mode
amplitudes and mode phases), while the bottom panel shows the actual data.

It is interesting to note that some of the signal peaks in the simulations
have been strengthened significantly by constructive interference with
noise peaks.  For example, a signal peak of 8\,ppm which happens to be in
phase with a 2$\sigma$ noise peak ($2\times4.7$\,ppm) will produce a peak
in power of 300\,ppm$^2$.  This illustrates the point made by
\citeone[Appendix A.2]{K+B95}: the effects of noise must be taken into
account when estimating the amplitude of a signal.

In these five simulations we can see that a signal of 8\,ppm is sometimes
obvious but sometimes not.  To quantify this, we have looked at the eight
strongest peaks in each power spectrum in the frequency range
1600--3000\,\muHz.  These are marked by vertical lines in \figref{fig.sim}.
Dotted lines show peaks that coincide with input frequencies or their 1/day
aliases to within $\pm 1.3\,\muHz$ (which is twice the rms scatter on the
differences between input and measured frequencies).  The remaining peaks,
marked by dashed lines, are assumed to be due to noise.  The results show
that 2--5 of the eight strongest peaks coincided with input frequencies and
a further 0--2 were 1/day aliases (the mean values for these were 4.0 and
1.0, respectively).

\if\preprint1
\begin{figure} 
\centerline{
\includegraphics[width=8cm, bb=68 59 351 648]
 {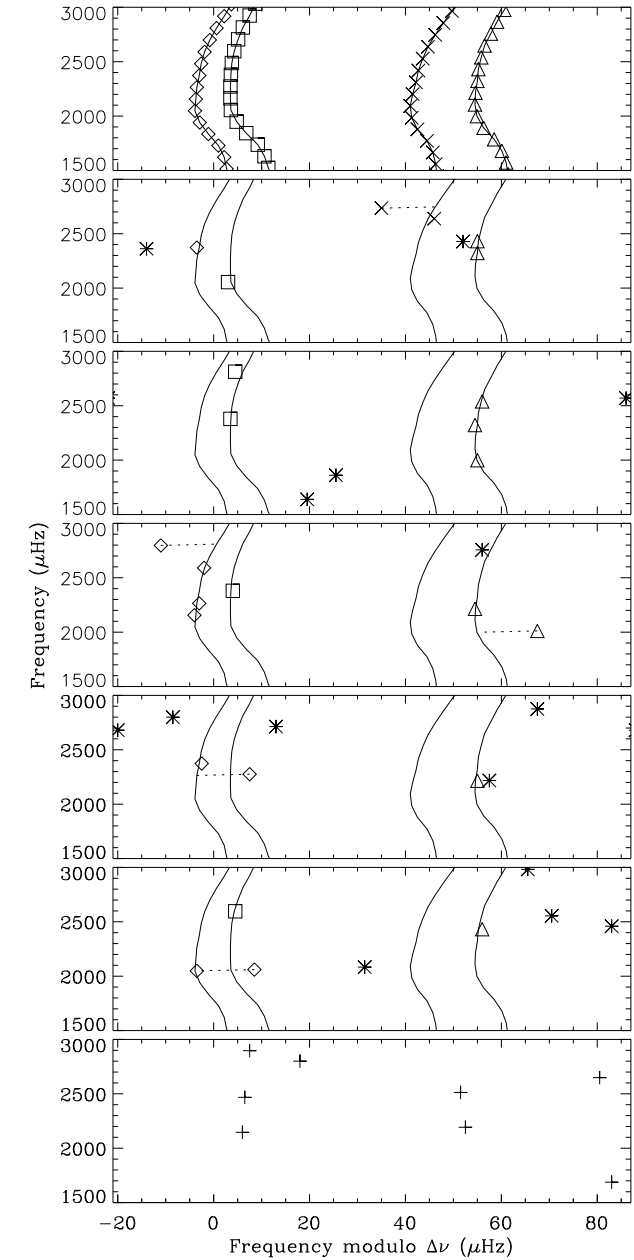}
}
\caption[]{\label{fig.echelle} Echelle diagrams based on \figref{fig.sim}.
The top panel shows the input model frequencies, plotted using $\Dnu{} =
108.0$\,\muHz.  The symbols indicate different $\ell$ values: 0 (squares), 1
(triangles), 2 (diamonds) and 3 (crosses).  The next five panels show the
eight strongest peaks in each simulation.  The solid lines show the locus
of input model frequencies.  Each peak is identified either as noise
(asterisks) or as one of the input frequencies.  In the latter case, some
peaks are shifted by 1/day (11.57\,\muHz), as indicated by dotted lines.
The bottom panel shows the eight strongest peaks in the actual data, with
$\Dnu{} = 107.0$\,\muHz.  }
\end{figure}

\fi

The same information is shown in \figref{fig.echelle} in the form of
echelle diagrams.  The top panel shows all input frequencies from
\citeone{ECD92}, although most of these are given very small amplitudes in
the simulations.  The symbols indicate different $\ell$ values: 0
(squares), 1 (triangles), 2 (diamonds) and 3 (crosses).  The next five
panels show the eight strongest peaks in each simulation.  The solid lines
show the loci of input model frequencies.  Each peak in \figref{fig.sim} is
identified either as noise (asterisks) or as one of the input frequencies.
In the latter case, some peaks are shifted by 1/day (11.57\,\muHz), as
indicated by dotted lines.

\subsection{Mode identifications}

Given that some of the eight strongest peaks in our power spectrum of
\aCenA{} may be real, we can attempt to identify the modes to which they
correspond.  Mode frequencies for low-degree and high-order oscillations in
the Sun and other solar-like stars are well approximated by the asymptotic
relation:
\begin{equation}
  \nu(n,\ell) = \Dnu{} (n + \half\ell + \epsilon) - \ell(\ell+1) D_0.
        \label{eq.asymptotic}
\end{equation}
Here $n$ and $\ell$ are integers which define the radial order and angular
degree of the mode, respectively; $\Dnu{}$ (the so-called large separation)
reflects the average stellar density, $D_0$ is sensitive to the sound speed
near the core and $\epsilon$ is sensitive to the surface layers.  Values
for these three parameters in the Sun are:
\begin{eqnarray*}
 \Dnu{}   & =  & 135.12 \pm 0.18\,\muHz \\
 D_0      & =  &   1.50 \pm 0.03\,\muHz \\  
 \epsilon & =  &   1.46 \pm 0.03
\end{eqnarray*}
We obtained these values by fitting \eqref{eq.asymptotic} to solar
frequency measurements for $n=17$--$25$ and $\ell = 0$--$2$ by
\citeone{FAA97}. 

The curvature in the \citeone{ECD92} frequencies for \aCenA{}, obvious in
\figref{fig.echelle}, indicates a departure from the asymptotic theory.
Curvature is also predicted by models of the Sun, at a level larger than
actually observed, reflecting the difficulty in modelling the solar
surface.  Therefore, the real oscillation frequencies in \aCenA{} are also
likely to show less curvature than the model by \citename{ECD92} This does
not affect the conclusions of the simulations, but is important when we try
to assign modes to our observed frequencies.

\if\preprint1
\begin{table*}
\caption[]{\label{tab.freqs} Frequencies of the eight strongest peaks in
the \aCen{} power spectrum and possible mode identifications $(n,\ell)$.}
\begin{tabular}{lllll}
\hline
\noalign{\smallskip}
Frequency~~~    &                         & \multicolumn{2}{c}{Possible identifications}           &                         \\
(\muHz)         &   Case 1                &      Case 2             &      Case 3              &      Case 4             \\
\noalign{\smallskip}                                                                             
\hline                                                                                           
\noalign{\smallskip}                                                                             
$1687.79$       & $\nu(13,3) -1/d\rlap?$       & noise                   & $\nu(15,1)      $        & $\nu(14,2)      $       \\
$2145.76$       & $\nu(18,1)      $       & $\nu(18,2)            $ & $\nu(20,0)      $        & $\nu(19,1)      $       \\
$2192.55$       & $\nu(18,2)      $       & $\nu(18,3)            $ & $\nu(20,1)      $        & $\nu(19,2)      $       \\
$2467.68$       & $\nu(21,1)      $       & $\nu(21,2)            $ & noise                    & noise                   \\
$2512.35$       & $\nu(21,2)      $       & $\nu(21,3)            $ & noise                    & $\nu(23,0) +1/d\rlap?$       \\
$2648.48$       & $\nu(22,3) -1/d\rlap?$       & noise                   & $\nu(25,0)      $        & $\nu(24,1)      $       \\
$2799.92$       & $\nu(24,1) +1/d\rlap?$       & $\nu(24,2) +1/d\rlap?      $ & noise                    & noise                   \\
$2896.28$       & $\nu(25,1)      $       & $\nu(25,2)            $ & $\nu(27,1)      $        & $\nu(26,2)      $       \\    
\noalign{\medskip}                                                                         
~~~$\Dnu{}/\muHz$ &\hfill$106.94 \pm 0.13$ & \hfill$106.99 \pm 0.16$ & \hfill$100.77 \pm 0.17$  & \hfill$100.77 \pm 0.17$ \\
~~~$D_0/\muHz$    &\hfill$  2.05 \pm 0.05$ & \hfill$  1.35 \pm 0.09$ & \hfill$  1.95 \pm 0.29$  & \hfill$  1.07 \pm 0.09$ \\
~~~$\epsilon$     &\hfill$  1.61 \pm 0.03$ & \hfill$  1.14 \pm 0.04$ & \hfill$  1.29 \pm 0.04$  & \hfill$  1.81 \pm 0.04$ \\
\noalign{\smallskip}
\hline
\end{tabular}
\end{table*}

\fi

The bottom panel of \figref{fig.sim} shows the eight strongest peaks in the
actual data.  Assuming our representation of the mode amplitudes in
\aCenA{} is realistic, we would expect some of these to correspond to
actual frequencies.  We have therefore searched for values of $\Dnu{}$
which fit the observed frequencies, as follows.  We calculated all pairwise
differences and looked for a value of $\Dnu{}$ which agreed with as many of
these as possible (within $\pm0.8\,\muHz$).  The only two values of
$\Dnu{}$ in the range 96--116\,\muHz{} which gave four or more coincidences
were 100.8\,\muHz{} (6 pairs) and 107.0\,\muHz{} (5 pairs).  We also
checked how well randomly chosen frequencies gave similar coincidences.  On
the basis of 25 simulations, we found that 6 coincidences occurred only
once.

For both of the values of $\Dnu{}$ found above, we attempted to identify
each of the eight observed frequencies by fitting to \eqref{eq.asymptotic}.
There are two possible identifications for each value of $\Dnu{}$ that have
$D_0$ in a physically realistic range (0.7--3.0\,\muHz{};
\citebare{ChD93b}) and these are given in \tabref{tab.freqs}.  We chose $n$
such that $\epsilon$ lay in the range 1--2.

We stress that we do not claim to have detected oscillations but that, if
we have, one of the four cases in the table represents the most likely
description.  In that case, the implied amplitude is 7--8\,ppm (1.3 times
solar).  If none of the eight frequencies is real, the oscillations must be
below 7\,ppm (1.2 times solar).  Either way, these observations represent
the most sensitive search yet made for stellar variability.




\if\preprint1
\begin{figure*}
 \includegraphics[
width=0.9\textwidth, bb=34 34 580 420] 
{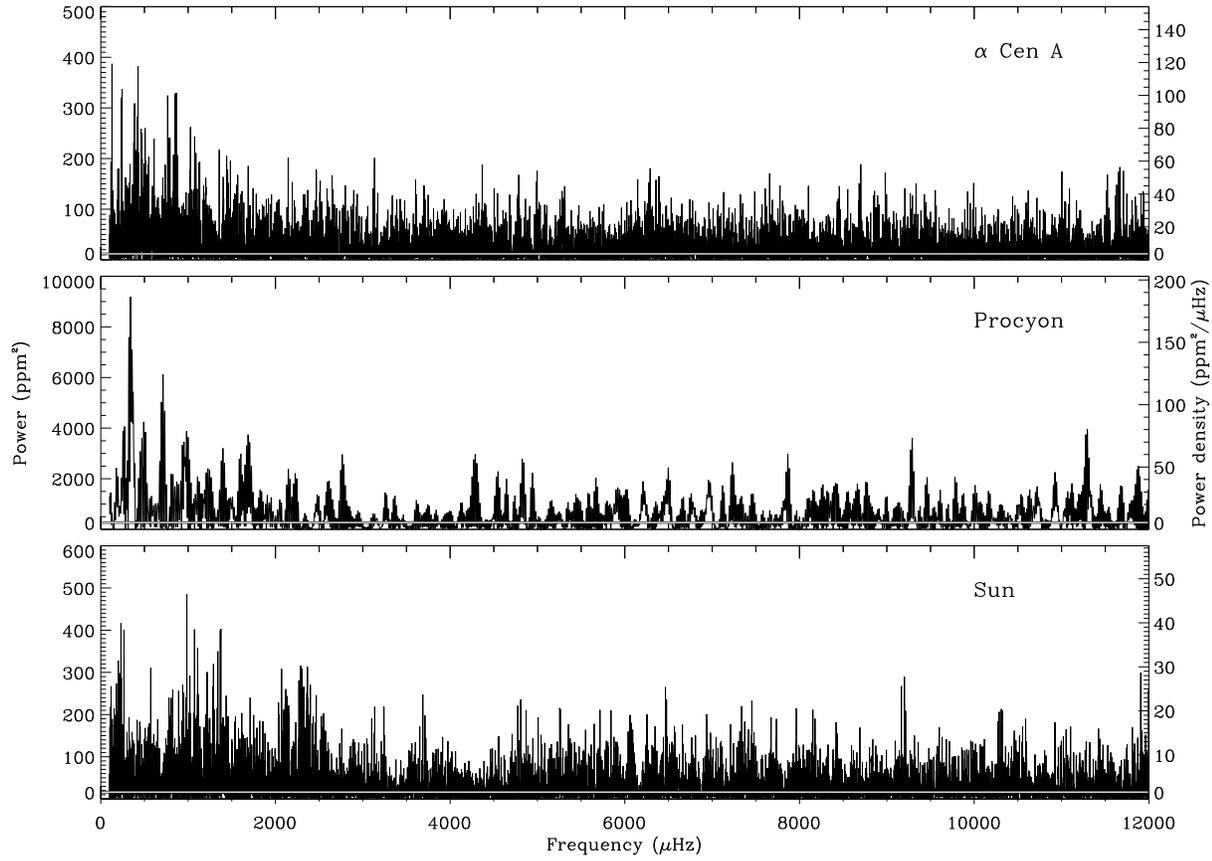}

\caption[]{\label{fig.power}Power spectra of equivalent-width observations
of \aCenA, Procyon and the Sun.  In each case, the mean level of white
noise is shown by the horizontal white line.  }
\end{figure*}

\fi

\section{Granulation power}	\label{sec.granulation}

The power spectrum of full-disk intensity measurements of the Sun shows a
sloping background due to granulation and other surface structure
\citeeg{RSRCJ97}.  This power background reflects surface temperature
fluctuations, so we expect it also to be present in EW measurements.
Indeed, our power spectrum of \aCenA{} (\figref{fig.power}) shows an excess
at low frequencies.  The same is true for our observations of Procyon and
the Sun, although there are much less data.

Note that the power density scales on the right-hand axes were obtained by
multiplying the left-hand scales (defined in \secref{sec.results}) by the
effective timespan of the observations (see Appendix A.1 of
\citebare{K+B95}).  For \aCenA, for example, the observations spanned
145\,hr from start to finish.  However, the measurements were given
different weights, so we calculated the effective timespan by integrating
under the weighted spectral window (right panel in \figref{fig.acen}); the
result was 85.5\,hr.  It is important to note that published power density
spectra are often calculated using different versions of Parseval's theorem
and must be multiplied by either two or four to be compared with our
definition of power density.

To investigate whether the low-frequency power excess observed in \aCenA{}
could arise from stellar granulation, we must make two corrections.
Firstly, we must subtract the contribution from white noise (photon noise,
instrumental and atmospheric effects), which we measure to have a level of
$7.3 \pm 0.2$\,ppm$^2$/\muHz.  This is shown by the horizontal line in
\figref{fig.power}.  Secondly, at frequencies below about 800\,\muHz{}
there is a deficit of power due to the high-pass filter (see
\secref{sec.filter}).

\if\preprint1
\begin{figure*}
\centerline{ \includegraphics[width=13cm]
{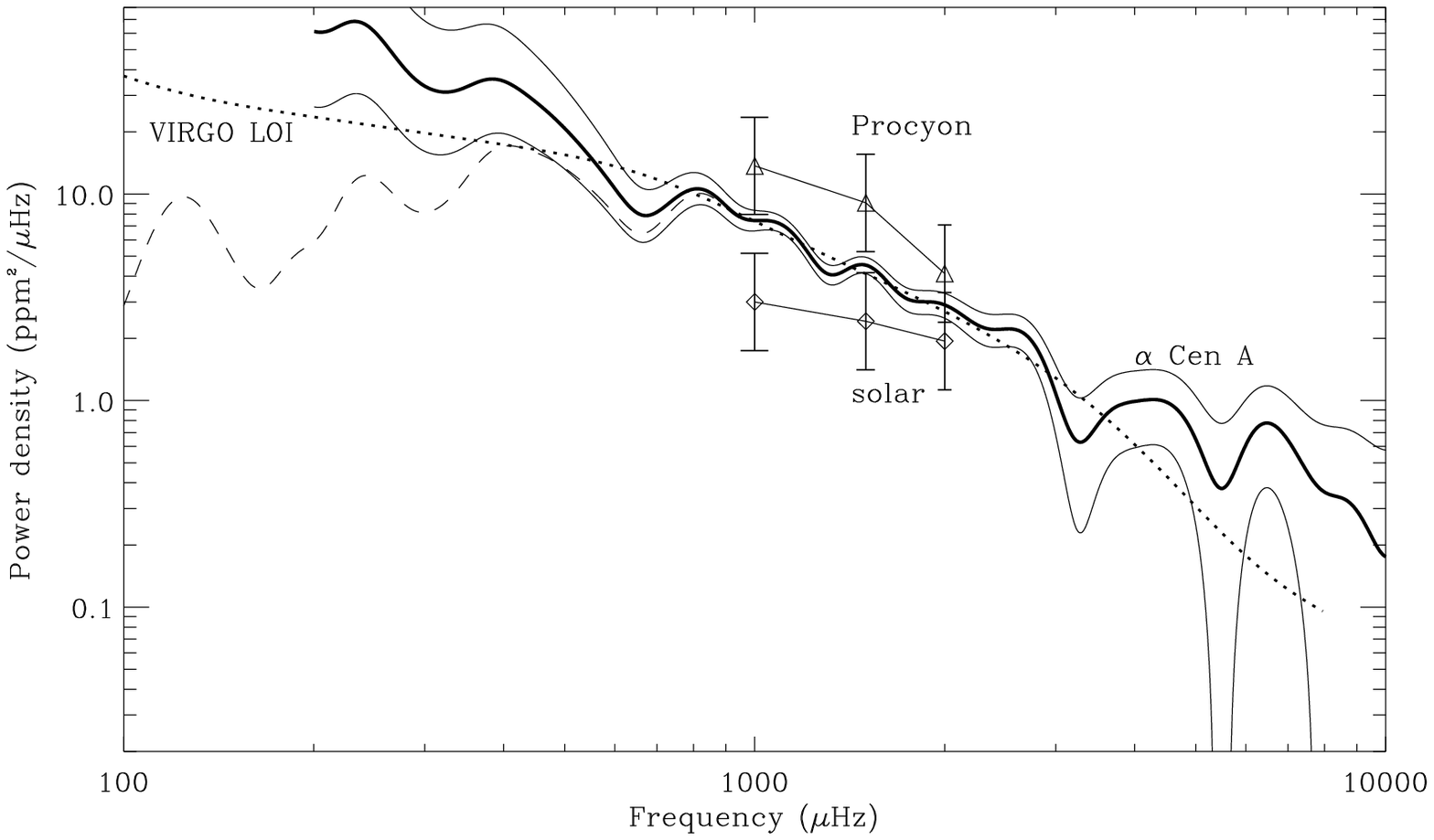}}
\caption[]{\label{fig.gran} Comparison of power density spectra.  The thick
solid curve shows the observed EW power density from our AAT/ESO
observations of \aCenA, after subtraction of white noise and after
correction at low frequencies for the effect of the high-pass filter.  The
two thinner solid lines on either side show the $\pm$2$\sigma$ errors in
these two corrections.  Dashed curve: same as thick solid curve, but
without correction for the high-pass filter.
The symbols show EW measurements from our AAT/ESO observations for Procyon
and the Sun, with 2$\sigma$ error bars.  The dotted line shows our estimate
of the EW granulation power density in the Sun, scaled from published
intensity measurements by the VIRGO LOI instrument (see text). }
\end{figure*}

\fi

The thick solid line in \figref{fig.gran} shows the power density from our
EW measurements of \aCenA{} after subtraction of the constant white noise
term, correction for high-pass filtering and smoothing.  The two thinner
solid lines on either side show the $\pm$2$\sigma$ errors in the two
corrections.

\if\preprint1
\begin{figure*}
 \includegraphics[
draft=false,
width=15cm, bb=40 293 563 461] 
{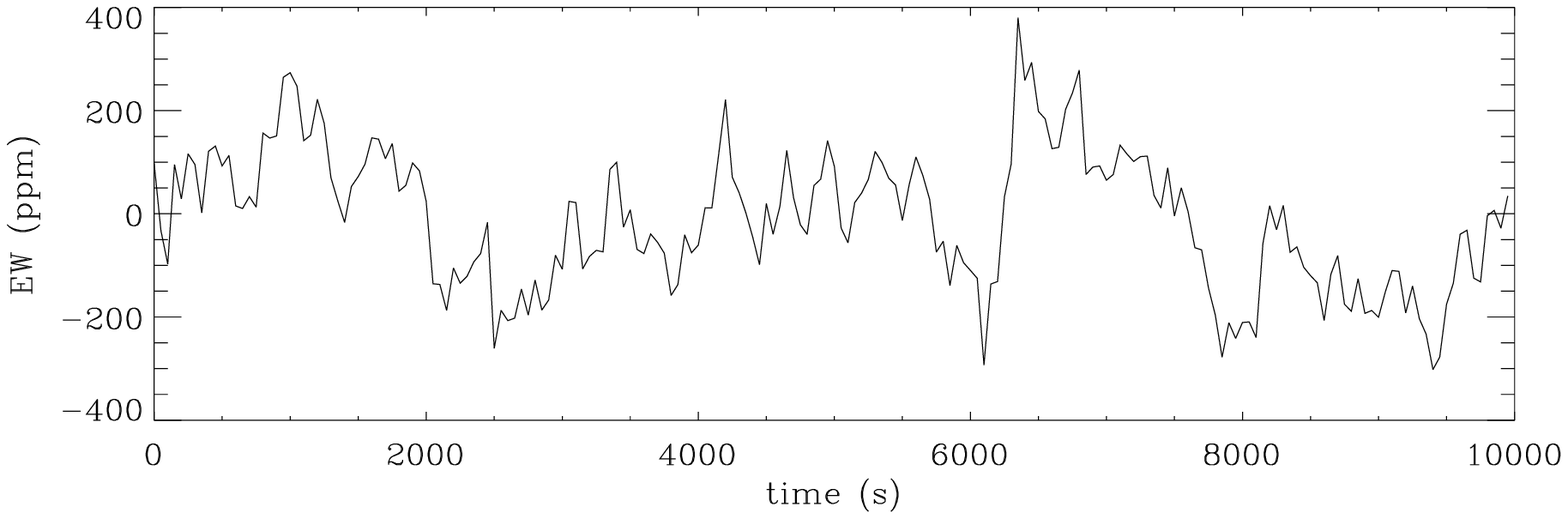}

\caption[]{\label{fig.simew} Example time series for EW measurements of
\aCenA{} based of the non-white component of the power spectrum. }
\end{figure*}

\fi

The power is quite uncertain at low frequencies due to uncertainties in the
removal of the effects of the high-pass filter, and at high frequencies due
to uncertainties in the subtraction of white noise component.  At
intermediate frequencies (600--3000\,\muHz), our measurement should be
accurate and in this regime the power is linear and well described by the
following relation:
\begin{equation}
  \log \mbox{(power density)} = G + P \log\left(\frac{\nu}{1\,\mbox{mHz}}\right)
        \label{eq.gran}
\end{equation}
where
\begin{eqnarray*}
 G & = &  0.85 \pm 0.09 \\
 P & = & -1.46 \pm 0.15
\end{eqnarray*}
and the power density is measured in ppm$^2$/\muHz.  The slope of the
background power is very similar to that produced by solar granulation
\citeeg{RSRCJ97}.  To compare the amount of power in our EW measurements
with published measurements of the total solar intensity, we must estimate
the conversion factor between the two observing methods.  For the Sun, a
fractional change in intensity of 1\,ppm that is caused by a temperature
change corresponds to a change in EW of the Balmer lines of 1.5\,ppm
\cite{BKR96}.  We assume the same conversion factor applies to temperature
fluctuations from granulation.  Furthermore, the amount of limb darkening
in the EW of Balmer lines -- and hence the signal from granulation -- is
greater than in total intensity \cite{BKR96} by a factor which we estimate
to be 1.15.  Hence, published measurements of the power density of solar
intensity should be multiplied by $(1.5)^2 (1.15)^2 = 3.0$.

We have applied this conversion to the recent measurement of full-disk
solar intensity measurements made with the LOI (Luminosity Oscillations
Imager) part of the VIRGO instrument on the SOHO spacecraft (Fig.~2 of
\citebare{AAF97}).  The result is shown as the dotted line in
\figref{fig.gran}.  Note that it was also necessary to multiply the LOI
values by four to bring them into line with our definition of power density
(T. Appourchaux, private communication).  The agreement with our
observations of \aCenA{} is excellent, giving strong evidence that we have
detected stellar granulation.

Our observations of Procyon and the daytime sky also produced power spectra
with excesses at low frequencies (\figref{fig.power}).  These power levels
are shown in \figref{fig.gran}, with 2$\sigma$ error bars.  Our solar
measurement is in reasonable agreement with the LOI data, although slightly
too low.  The granulation power in Procyon appears to be greater than that
in \aCenA{} by a factor of 2.0, although this result is only at the
2$\sigma$ level.

To illustrate the size of the granulation signal from \aCenA, we show in
\figref{fig.simew} an example time series, sampled at 50\,s intervals,
based on the non-white component of the power spectrum (the thick line in
\figref{fig.gran}).  This is how the EW signal from \aCen{} would look
without photon noise or measurement errors.  

When one is trying to detect oscillations, these fluctuations in EW due to
stellar granulation represent a fundamental noise source.  For example,
suppose that the weather had been 100\% clear and that all of our EW
measurements (H$\alpha$ and H$\beta$) were photon-noise limited.  Then we
would have expected a white noise level in the amplitude spectrum (from
photon noise) of 1.4\,ppm and a noise level from granulation at 2.3\,mHz of
1.9\,ppm.  The total noise at 2.3\,mHz would then have been 2.4\,ppm, about
half the value that we actually measured.  In other words, we were a factor
of two away from the limit set by photon and granulation noise, presumably
due to instrumental and atmospheric effects.  Had we achieved this limit, a
signal with solar amplitude (about 6\,ppm) would have been detected.

Given a CCD with sufficiently fast readout, it would be possible to record
a much larger part of the spectrum and hence to measure EWs of many more
lines at a higher duty cycle \cite{BKR96}.  The photon noise would then be
reduced by up to a factor of about two, to which granulation noise must
still be added.  This represents the fundamental limit for the EW method.

Granulation is also a fundamental limit for intensity and colour
measurements, but much less critically for Doppler shift measurements,
since the background power in the solar velocity spectrum is a tiny
fraction of the oscillation amplitudes \cite{GCG97}.



\section{Conclusions}


Our observations of \aCenA{} represent the most sensitive search yet made
for solar-like oscillations.  We can set a firm upper limit on the
amplitudes of 1.4 times solar, which is approximately the level at which
oscillations are expected.  We find tentative evidence for $p$ modes in the
form of a regular series of peaks.  If these do not correspond to real
oscillations then the upper limit on amplitude becomes 1.2 times solar.

We find an excess of power at low frequencies in \aCenA{} which has the
same slope and strength as power from granulation in the Sun.  We therefore
suggest that we have made the first detection of granulation power in a
solar-like star.

\subsection*{Acknowledgements}

The observations would have been impossible without the excellent support
we received from staff at both observatories.  We are especially grateful
to Roy Antaw, Bob Dean, Sean Ryan, John Stevenson and Gordon Shafer at the
AAO and to Luca Pasquini, Peter Sinclaire and Nicolas Haddad at ESO\@.  We
also thank both committees (ATAC and OPC) for allocating telescope time and
the AAO Director for granting the sixth AAT night.  This work was supported
financially by the Australian Research Council, by the Danish Natural
Science Research Council and by the Danish National Research Foundation
through its establishment of the Theoretical Astrophysics Center.

\if\preprint0
        \clearpage
        \renewcommand{\baselinestretch}{1}
	
        \renewcommand{\baselinestretch}{2}
        \clearpage
	\begin{figure*} 

\includegraphics[
width=0.9\textwidth, bb= 90 147 524 394]
 {figures/weights.ps}
\caption[]{\label{fig.weights} The final time series of EW observations of
\aCenA.  The upper panel shows the square root of statistical weights (the
inverse of the local rms scatter) in arbitrary units for all AAT
measurements (both H$\alpha$ and H$\beta$ are shown).  There are 8430
points, with 5773 having non-zero weight.  For points with zero weight, we
have added a random vertical scatter to make it easier to see their
distibution in time.  The middle panel shows the same for ESO data
(H$\alpha$ only); there are 8847 points, with 7999 having non-zero weight.
The bottom panel shows all EW measurements having non-zero weight (13772
points).  The peak-to-peak height of the individual modes we are trying to
detect corresponds approximately to the size of the dots in the figure.  }
\end{figure*}

  
	\begin{figure*}
 \includegraphics[draft=false,width=15cm, bb=49 289 563 461] 
{figures/acen-power.ps}

\caption[]{\label{fig.acen}Power spectrum of equivalent-width observations
of \aCenA{} in the region where signal would be expected.  The right-hand
plot shows the spectral window, with the same horizontal scale, calculated
by taking the power spectrum of a sinusoid with an amplitude of 1\,ppm and
the same sampling and weights as the actual data. }
\end{figure*}

	\begin{figure*} 

\includegraphics[width=13cm, bb=74 56 503 648]
 {figures/newsim.ps}
\caption[]{\label{fig.sim}Simulated power spectra for \aCenA, using the
same sampling times and data weights as the actual observations.  The top
panel shows a simulation without noise, while the next five panels show
simulations with noise.  The eight strongest peaks are marked by vertical
lines, with dotted lines indicating those coinciding with input frequencies
or their 1/day aliases, and dashed lines showing noise peaks (see also
\figref{fig.echelle}).  The bottom panel shows the actual data from \aCenA,
with the eight strongest peaks being marked (see \tabref{tab.freqs}).  }
\end{figure*}

	\begin{figure} 
\centerline{
\includegraphics[width=8cm, bb=68 59 351 648]
 {figures/echelle.ps}
}
\caption[]{\label{fig.echelle} Echelle diagrams based on \figref{fig.sim}.
The top panel shows the input model frequencies, plotted using $\Dnu{} =
108.0$\,\muHz.  The symbols indicate different $\ell$ values: 0 (squares), 1
(triangles), 2 (diamonds) and 3 (crosses).  The next five panels show the
eight strongest peaks in each simulation.  The solid lines show the locus
of input model frequencies.  Each peak is identified either as noise
(asterisks) or as one of the input frequencies.  In the latter case, some
peaks are shifted by 1/day (11.57\,\muHz), as indicated by dotted lines.
The bottom panel shows the eight strongest peaks in the actual data, with
$\Dnu{} = 107.0$\,\muHz.  }
\end{figure}

	\begin{figure*}
 \includegraphics[
width=0.9\textwidth, bb=34 34 580 420] 
{figures/power.ps}

\caption[]{\label{fig.power}Power spectra of equivalent-width observations
of \aCenA, Procyon and the Sun.  In each case, the mean level of white
noise is shown by the horizontal white line.  }
\end{figure*}

	\begin{figure*}
\centerline{ \includegraphics[width=13cm]
{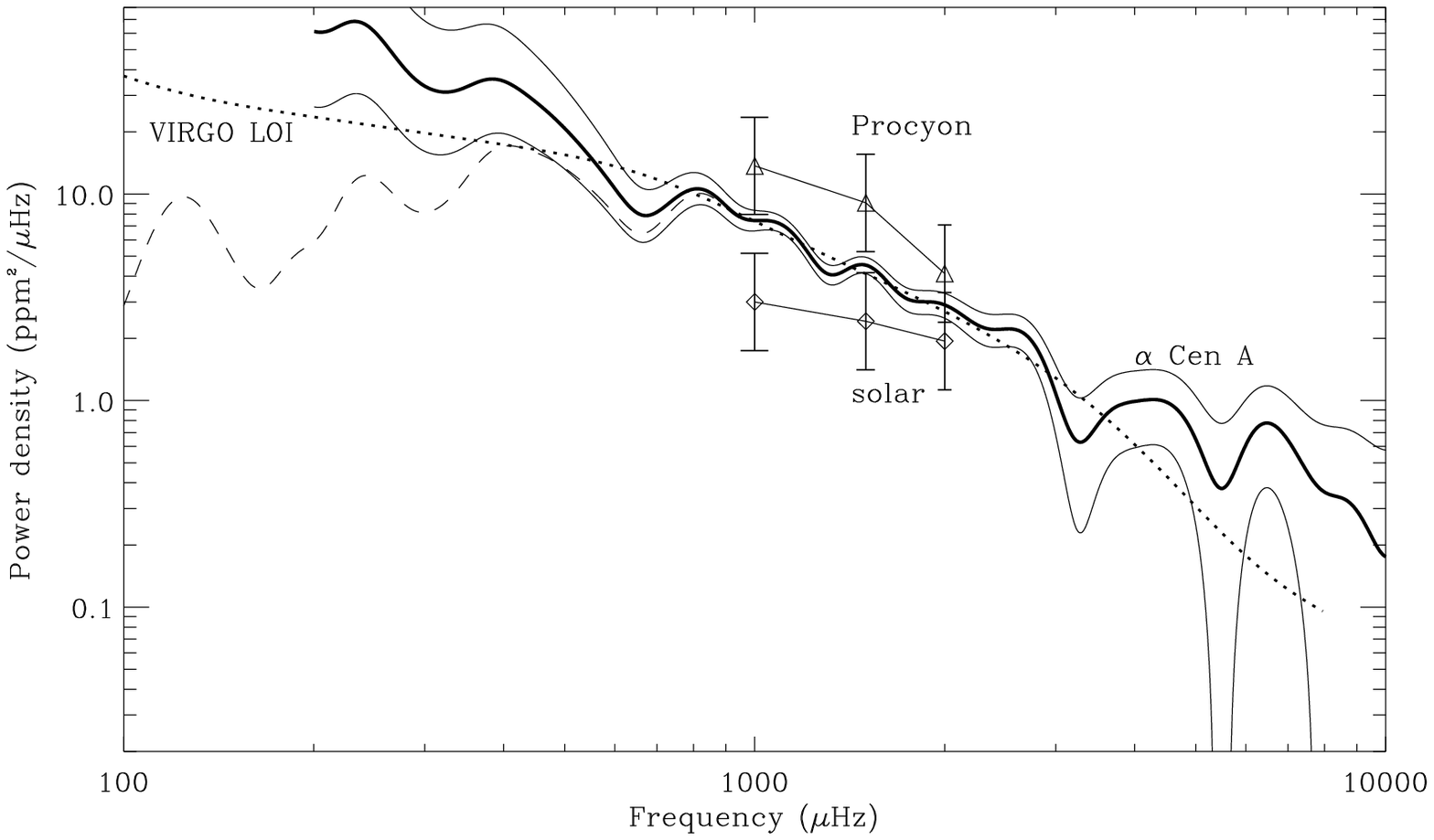}}
\caption[]{\label{fig.gran} Comparison of power density spectra.  The thick
solid curve shows the observed EW power density from our AAT/ESO
observations of \aCenA, after subtraction of white noise and after
correction at low frequencies for the effect of the high-pass filter.  The
two thinner solid lines on either side show the $\pm$2$\sigma$ errors in
these two corrections.  Dashed curve: same as thick solid curve, but
without correction for the high-pass filter.
The symbols show EW measurements from our AAT/ESO observations for Procyon
and the Sun, with 2$\sigma$ error bars.  The dotted line shows our estimate
of the EW granulation power density in the Sun, scaled from published
intensity measurements by the VIRGO LOI instrument (see text). }
\end{figure*}

	\begin{figure*}
 \includegraphics[
draft=false,
width=15cm, bb=40 293 563 461] 
{figures/simew.ps}

\caption[]{\label{fig.simew} Example time series for EW measurements of
\aCenA{} based of the non-white component of the power spectrum. }
\end{figure*}

\else
        \bsp
\fi

\end{document}